# Demonstration of a Degenerate Band Edge in Periodically-Loaded Circular Waveguides

Mohamed A. K. Othman, *Student Member, IEEE* and Filippo Capolino, *Senior Member, IEEE*

*Abstract*—We demonstrate the existence of a special degeneracy condition, called degenerate band edge (DBE), between two Bloch modes in periodically loaded circular all-metallic waveguides at microwave frequencies. The DBE condition has been associated with a dramatic reduction in group velocity and with some unique resonance properties, but it has not been shown in hollow waveguide structures yet. Hence, we show here its existence in two periodic waveguide examples. The unit cell of the first case is composed of a circular waveguide loaded with two inner cylinders with elliptical irises with misaligned angles. The second structure is composed of loading the waveguide with elliptical rings. The demonstration of DBE in those waveguide is explained through a simple multi transmission line approach where the conditions to obtain DBE are clarified, and suggests that the DBE can occur in several other analogous periodic waveguide. These structures can be potentially used to investigate unconventional gain schemes in traveling wave tubes or other kinds of distributed amplifiers and novel pulse compressors.

*Index Terms*—slow wave structures, periodically-loaded waveguide, periodic structure, degenerate band edge.

## I. INTRODUCTION

Slow-wave structures (SWSs) like coupled-cavity traveling wave tubes, are one of the main building block in high power microwave electronic devices such as amplifiers and oscillators [1]. Among the various implementations of periodic SWSs, here we are investigating those that provide significant reduction in the group velocity of the electromagnetic propagating mode, which can result in new phenomenological improvements in resonator quality factors, reduction of size, and mitigation of losses [2], [3]. We focus here on a particular class of SWSs based on periodic structures with at least two Bloch modes that exhibit a degeneracy conditions, such as the degenerate band edge (DBE) condition. This condition has not been explored yet in waveguides like those in this paper. The DBE phenomena was first introduced in [3]–[5] by observing frozen mode regimes in multilayer anisotropic dielectric structures. In general, the DBE condition causes a quartic power dependence at the band-edge of the Bloch wavenumber $k$ versus frequency $\omega$, i.e.,

$$\omega - \omega_d = \alpha (k - k_d)^4, \qquad (1)$$

This research was supported by AFOSR MURI Grant FA9550-12-1-0489 administered through the University of New Mexico.

M. A. K. Othman and F. Capolino are with the Department of Electrical Engineering and Computer Science, University of California Irvine, Irvine, CA 92697 USA. (e-mail: mothman@uci.edu, f.capolino@uci.edu).

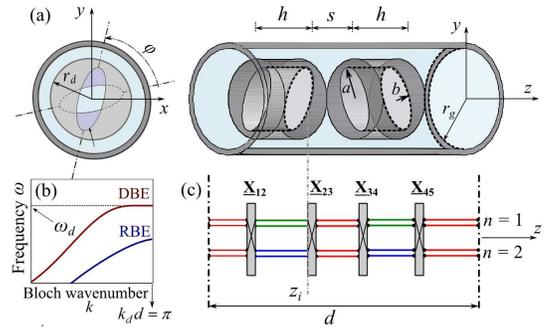

Fig. 1. (a) A unit cell composed of a circular air-filled waveguide loaded with two discs with elliptical irises misaligned by an angle $\varphi$. The periodic structure supports a DBE under certain conditions. (b) Typical dispersion relation of two modes one exhibiting an RBE and another exhibiting DBE. (c) Equivalent multi TL that is adopted to understand the occurrence of the two modes in (b) and DBE in particular for the structure (a).

where $\omega_d$ is the DBE angular frequency, $k_d = \pi/d$ represents the edge of the Brillouin zone, $d$ is the length of the periodic structure unit cell (Fig. 1), and $\alpha$ is a problem dependent constant. Consequently, not only the group velocity vanishes, i.e., $v_g = \partial_k \omega = 0$, but DBE implies also that the first and second derivative of the group velocity are identically zero, i.e., $\partial_k v_g = 0$ and $\partial_k^2 v_g = 0$, with the only the third derivative of the group velocity non-zero, $\partial_k^3 v_g \neq 0$ (the partial derivative is defined here as $\partial_k^n \omega \equiv \partial^n \omega / \partial k^n$). This leads to a gigantic increase in the density of electromagnetic modes (or density of states) and the group index as we have shown in [3]. Due to the extremely low group velocity of modes close to the DBE condition inside the structure, large field enhancement occurs [2]-[5], and structures exhibiting DBE modes could potentially be used as novel SWS in TWT for high power generation [6], by observing fundamental changes in the amplification mechanisms compared to those in conventional TWTs [1], [6]. Moreover, it has been shown that, in principle, gain in DBE-based laser cavities is largely enhanced compared to standard Fabry–Pérot cavities and even compared to cavities utilizing periodic structures based on a single mode operation [3]. At radio frequencies, DBE condition have been observed only in periodically-coupled microstrip lines [7] but not yet in metallic waveguides as in this paper, that can be utilized in a variety of high power devices. In this letter we propose for the first time a realization of a DBE in periodic metallic waveguide structures, such as the periodically loaded circular waveguide unit cells in Fig. 1.

## II. DEGENERATE BAND EDGE IN LOADED WAVEGUIDES

Consider the perfect electric conductor (PEC) air-filled circular waveguide with a radius $r_g$ and periodic unit cell of

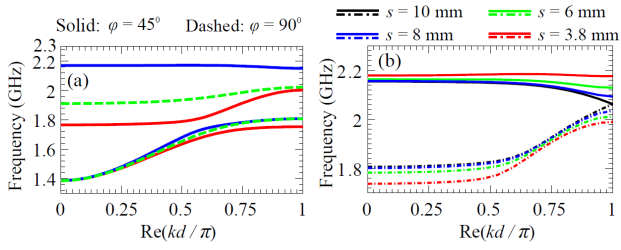

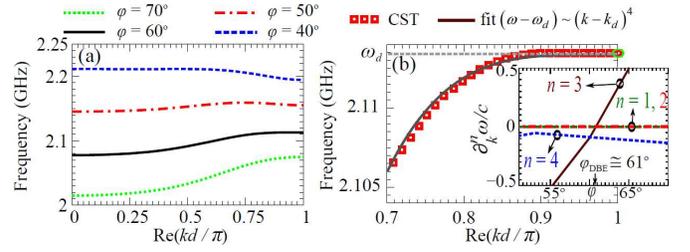

Fig. 2. (a) Dispersion diagram of the periodic structure in Fig. 1 for different misalignment angles of the elliptical irises for $h = 10$ mm and $s = 5$ mm. Each dashed line is split into two by changing $\varphi$ from 90 to 45degrees. (b) Same as (a), but for varying $s$ for $\varphi = 45°$, and we show only two higher order modes.

Fig. 3. (a) Dispersion diagram of the highest order mode in Fig. 2(b) obtained by varying the misalignment angle $\varphi$ with $s = 3.8$ mm. (b) Dispersion for $\varphi = 61°$ around 2.11 GHz and showing an accurate fit to equation (1), and the inset shows the normalized derivatives of group velocity at $k_d$ proving the existence of a DBE at $\varphi_{DBE} \cong 61°$.

length $d$, as in Fig. 1(a). The waveguide is loaded with two discs of equal length $h$, with an elliptic iris each, separated by a distance $s$. The two elliptical irises have identical dimensions and aspect ratio, i.e., the same major and minor radii $b$ and $a$, as shown in Fig. 1(a). However, their major axes are misaligned by an angle $\varphi$. Let $\mathbf{E}_t(\boldsymbol{\rho},z)$ and $\mathbf{H}_t(\boldsymbol{\rho},z)$ be the transverse components of the electric and magnetic fields at a cross-section plane defined by $z = $ constant. They can be represented as expansions of the normal modes inside the waveguide at any $z$ as $\mathbf{E}_t(\boldsymbol{\rho},z) = \sum_n \mathbf{e}_n(\boldsymbol{\rho}) V_n(z)$ and $\mathbf{H}_t(\boldsymbol{\rho},z) = \sum_n \mathbf{h}_n(\boldsymbol{\rho}) I_n(z)$, where $\mathbf{e}_n(\boldsymbol{\rho})$ and $\mathbf{h}_n(\boldsymbol{\rho})$ are the $n^{th}$ electric and magnetic eigenvectors that depend on the local transverse cross section [8]. Scalars $V_n$ and $I_n$, interpreted as transmission line's (TL) voltage and current, are the amplitudes of those fields. To treat the DBE condition it has been convenient to define a state vector $\boldsymbol{\Psi}(z) = \begin{bmatrix} \mathbf{V}^T(z) & \mathbf{I}^T(z) \end{bmatrix}^T$, that describes the evolution of electromagnetic waves along the $z$-direction using a TL approach[8]. At an interface between two different waveguide cross-sections (for example, in Fig. 1, at $z_i$ located just before the end of the first ring), we apply the boundary conditions matching the transverse fields in each of the two segments $\mathbf{E}_t(\boldsymbol{\rho},z_i) = \mathbf{E}_t(\boldsymbol{\rho},z_{i+1})$ and $\mathbf{H}_t(\boldsymbol{\rho},z_i) = \mathbf{H}_t(\boldsymbol{\rho},z_{i+1})$. Then we define an interface matrix $\underline{\mathbf{X}}_{i,i+1}$ that is composed of the projections of the electric and magnetic eigenmode functions belonging to both waveguide segments at $z_i$ and $z_{i+1}$ (see [8] for detailed analysis). Hence the state-vector is transformed as $\boldsymbol{\Psi}(z_i) = \underline{\mathbf{X}}_{i,i+1} \boldsymbol{\Psi}(z_{i+1})$ across a cross section discontinuity. The rotation matrix $\underline{\mathbf{X}}_{i,i+1}$ is strongly dependent on the misalignment angle $\varphi$ in Fig. 1, that mixes modes in the various segments and therefore has a significant impact on the dispersion of modes and the possibility of attaining different band edge condition, among which DBE is of great interest[4], [5], as to be shown later. A detailed formulation is developed in [6] using transfer matrices. Here, by using full-wave simulations, we demonstrate the DBE is obtained in a realistic waveguide structure, confirming the prediction based on TL formulation, for the periodic structure as in Fig. 1, with $r_g = 40$ mm, $r_d = 35$ mm, $b = 25$ mm, and $a = 10$ mm. In Fig. 2(a) we report the dispersion diagram of the Bloch mode supported by the periodic structure using the eigenmode solver by CST Microwave Studio for two different misalignment angles, $\varphi = 45°$ and $\varphi = 90°$. We only show the positive real branch of the diagram with normalized Bloch wavenumber

$kd/\pi$ along the $z$-direction, varying frequency. We observe that for $\varphi = 90°$ (dashed lines in Fig. 2(a)) in the shown frequency range, the structure supports four modes (with positive $k$). In fact, each dashed curve in Fig. 2(a) corresponds to two curves on top of each other, and forms a couple of modes orthogonally-polarized, with a 90-degree rotation symmetry in the $x$-$y$ plane obeying the same dispersion relation. All four modes exhibit regular band edges (RBE), just below the cutoff frequency of the hollow circular waveguide (~2.197GHz). Indeed, if we consider the equivalent periodic multi-TL model in Fig. 1(b), with at least two uncoupled TL segments and mode coupling matrices $\underline{\mathbf{X}}$ at each cross section discontinuity that introduces mixing between TL modes, the mixing associated to 90 degrees misalignment is not sufficient to develop a DBE. However, when $\varphi = 45$ degrees, we report that each dashed line Fig. 2(a) splits into two different curves, which means that now $\underline{\mathbf{X}}$ introduces significant mixing between the two TL modes that breaks symmetry between the two polarizations. We then vary the separation $s$ between the rings, and in Fig. 2(b) we show the dispersion diagram for the two higher order modes in Fig. 2(a), with $\varphi = 45°$, for different $s$. For smaller separation such as $s = 3.8$ mm, a DBE condition may be qualitatively observed, relative to the higher order mode that becomes degenerate with the lower order mode at the edge of the Brillouin zone, i.e., it has $\mathrm{Re}(k) = k_d$, and this is suggested because of the flatness of the dispersion curve (solid red curve in Fig. 2(b)) relative to other cases with larger $s$. This is the special degeneracy we discuss in this paper. However the mathematical degeneracy condition is yet to be precisely found; that dictates vanishing group velocity, its first as well as second derivatives at $k_d$. To find out when this feature exactly arises, we vary the rotation angle $\varphi$ and plot the dispersion diagram in Fig. 3(a) for the case with $s = 3.8$ mm shown in Fig. 2(b). We can see that the band edge feature for the highest order mode (dispersion near $k_d$) is prone to the variation of $\varphi$ whereas we have observed that the lower order mode in Fig. 2(b) is not sensitive to the same variation (not shown here). This indicates the possibility of acquiring DBE for the mode in Fig. 3(a) by optimizing $\varphi$. We plot the derivatives $\partial_k^n \omega$ with $n \in \{1,2,3,4\}$ we observe that for $\varphi \cong 61°$ the two derivatives of group velocity vanish while the third derivative is non zero and negative; nonetheless group velocity and its first derivative are still vanishing for wide range of which indicates that the dispersion can still be maintained relatively flat near $k_d$ (see Fig. 3(a) for angles 50 and 65 degrees). For the case with $\varphi = \varphi_{DBE} = 61°$, we

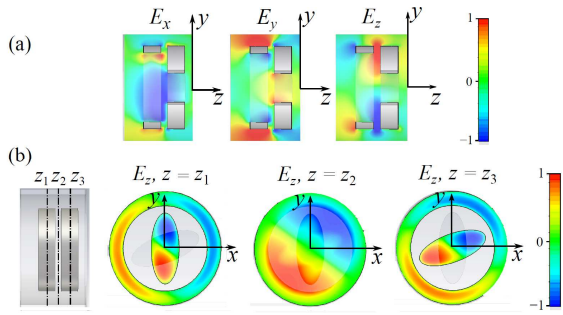

Fig. 4. (a) Normalized electric field distribution (three components, each normalized to its maximum) in the longitudinal plane of the mode that exhibit a DBE for the structure in Fig. 1. (b) The normalized z-directed electric field along the +z-direction in the transverse plane at three different locations inside the unit cell.

confirm that the dispersion develops a DBE by observing how full wave dispersion well approximates equation (1) near the band edge by means of numerical fitting. The parameters obtained from numerical fitting are $\omega_d \cong 2\pi(2.114\,\text{GHz})$ and $\alpha \cong -4.52\times 10^3$ (units of $\text{m}^4\text{s}^{-1}$) a problem dependent constant with $\alpha<0$. It is in general proportional to the non-vanishing fourth derivative of $\omega$ with respect to $k$, at $k_d$ [4], hence it can be obtained from the inset of Fig 3(b) as $\alpha = \partial_k^4 \omega/24 \cong -4.53\times 10^3$ ($\text{m}^4\text{s}^{-1}$). The fitting parameters in this model are found by setting a root mean squared error (RMSE) in the order of $\sim 10^{-5}$, which indicates a very good fitting. In Fig. 4 we show the normalized electric field distribution of the eigenmode that exhibits a DBE at $\omega_d$. The three components of the modal electric field are plotted in Fig. 4(a) in the y-z plane at $x = 0$, each component is normalized to its own maximum. We report the DBE mode electric field distribution with a strong axial component (simulation data show negligible longitudinal magnetic field $H_z$ everywhere inside the waveguide, not shown here for brevity, indicating a transverse magnetic (TM) like mode at DBE). Also we plot the longitudinal electric field at different x-y transverse cut-planes inside the structure in Fig. 4(b) showing the field inside irises as well as between the discs. Analogous optimized TM structures, may be utilized for electron beam devices in high power generation [6], since DBE mode in this structure is a slow-wave mode with a phase velocity of $\omega/k_d \sim 0.5c$ that can be synchronized to an electron beam with optimized shape [1], [6]. Following the same procedure outlined above, in Fig. 5(a) we show a somehow simpler geometry with two hollow elliptical rings per unit cell. The rings' major axes are misaligned by an angle $\varphi$. For $\varphi = 90$ degrees, we report the lowest order modes in the structure and we observe as in the previous case, there are two degenerate but orthogonal polarization modes represented in the dispersion relation in Fig. 5(b) by dotted-dashed lines, exhibiting an RBE. However, for different misalignment angles, mode splitting is evident and DBE is observed for $\varphi= 53$ degrees, where we show the fitting to (1) with $\omega_d \cong 2\pi(1.731\,\text{GHz})$, $\alpha \cong -1.7\times 10^4$ ($\text{m}^4\text{s}^{-1}$) and with an RMSE of $\sim 3\times 10^{-5}$. The axial electric field distribution of the DBE at $k_d$ is shown in Fig. 5(c). We have also seen that severing each ring in Fig. 5 into thinner rings with careful design the periodic structure still does not impair the capability to achieve DBE, that would be beneficial for mitigating conductor losses. Other design considerations

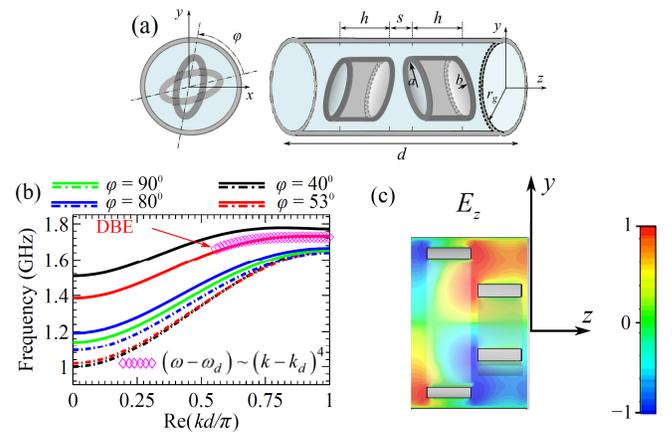

Fig. 5. Ring-loaded waveguide unit cell. (a) Dispersion diagram of the periodic structure for different misalignment angles of the elliptical rings, with $h$= 15 mm and $s$= 2.5 mm, $b$ = 25 mm, $a$ = 10 mm, and ring thickness is 5 mm. For $\varphi = 53°$, the dispersion develops a DBE showing a fitting with the DBE dispersion relation indicated by the symbols. (c) Normalized z-directed electric field in the y-z plane ($x$= 0) of the DBE mode.

including fixtures for the rings, impedance and loading effects will be studied in the future.

### III. CONCLUSION

We have demonstrated the existence of a special degeneracy condition, called degenerate band edge (DBE), between two modes in metallic circular waveguides. This is achieved by loading the waveguide with elliptical irises or elliptical rings with carefully designed misalignment angle. Full-wave eigenmode results confirmed the existence of DBE modes with an axial electric field component. The peculiar resonance condition associated to this special degeneracy condition and its unique advantages have been discussed previously [3−6] and they can be obtained now also in metallic waveguide technology. This allows us to explore potential applications such as the enhancement of wave-electron beam interaction in slow wave structures. In a similar contest we have already shown a framework for such degeneracy condition that provides superior gain conditions compared to other types of uniform or periodic structure gain media [3], thus our proposed structures promise a great impact on high power amplifier and oscillator designs.

### ACKNOWLEDGMENT

The authors thank CST Inc. for providing CST Microwave Studio that was instrumental in this work. The authors also would like to thank Prof. Alex Figotin at the University of California Irvine for fruitful and continuous discussions.

### REFERENCES

[1] A. S. Gilmour, *Principles of traveling wave tubes*. Artech House, 1994.
[2] A. Figotin and I. Vitebskiy, "Oblique frozen modes in periodic layered media," *Phys. Rev. E*, vol. 68, no. 3, p. 036609, Sep. 2003.
[3] M. Othman, F. Yazdi, A. Figotin, and F. Capolino, "Giant Gain Enhancement in Photonic Crystals with a Degenerate Band Edge," *arXiv:1411.0657*, Nov. 2014.
[4] A. Figotin and I. Vitebskiy, "Frozen light in photonic crystals with degenerate band edge," *Phys. Rev. E*, vol. 74, no. 6, p. 066613, Dec. 2006.
[5] A. Figotin and I. Vitebskiy, "Slow wave phenomena in photonic crystals," *Laser Photonics Rev.*, vol. 5, no. 2, pp. 201–213, Mar. 2011.
[6] M. Othman, V. A. Tamma, and F. Capolino, "Complex modes and new amplification regimes in periodic multi transmission lines interacting with an electron beam," *ArXiv Prepr. ArXiv14111046*, 2014.
[7] C. Löcker, K. Sertel, and J. L. Volakis, "Emulation of propagation in layered anisotropic media with equivalent coupled microstrip lines," *Microw. Wirel. Compon. Lett. IEEE*, vol. 16, no. 12, pp. 642–644, 2006.
[8] Felsen L. B. and Marcuvitz N., *Radiation and Scattering of Waves*, Wiley-IEEE Press. 1994.